\documentclass[aoas,MSNbibl,nameyear,dvips]{arximspdf}
\usepackage{dcolumn,url,breakurl}
\usepackage{graphicx}

\doi{10.1214/13-AOAS707} 
\volume{8}
\issue{2}
\pubyear{2014}
\firstpage{801}
\lastpage{823}

\makeatletter
\newcolumntype{d}[1]{D{.}{.}{#1}}
\renewcommand{\citep}[1]{(\citeauthor{#1} \citeyear{#1})}
\newcommand{\eqref}[1]{(\ref{#1})}
\makeatother

\begin{document}
\begin{frontmatter}

\title{Testing the disjunction hypothesis using Voronoi
diagrams with applications to genetics\thanksref{T1}}
\runtitle{Simultaneous inference with bivariate data}
\thankstext{T1}{Supported from the National Science Foundation Grant
(Award Number ABI-1262538).}

\begin{aug}
\author{\fnms{Daisy} \snm{Phillips}\corref{}\ead[label=e1]{dlp245@psu.edu}}
\and
\author{\fnms{Debashis} \snm{Ghosh}\ead[label=e2]{ghoshd@psu.edu}}
\affiliation{Pennsylvania State University}
\runauthor{D. Phillips and D. Ghosh}
\address{Department of Statistics\\
Pennsylvania State University\\
University Park, Pennsylvania 16802\\
USA\\
\printead{e1}\\
\phantom{E-mail:\ }\printead*{e2}}
\end{aug}

\received{\smonth{3} \syear{2012}}
\revised{\smonth{10} \syear{2013}}

%
\begin{abstract}
Testing of the disjunction hypothesis is appropriate when each gene or
location studied is associated with multiple $p$-values, each of which is
of individual interest. This can occur when more than one aspect of an
underlying process is measured. For example, cancer researchers may
hope to detect genes that are both differentially expressed on a
transcriptomic level and show evidence of copy number aberration.
Currently used methods of $p$-value combination for this setting are
overly conservative, resulting in very low power for detection. In this
work, we introduce a method to test the disjunction hypothesis by using
cumulative areas from the Voronoi diagram of two-dimensional vectors of
$p$-values. Our method offers much improved power over existing methods,
even in challenging situations, while maintaining appropriate error
control. We apply the approach to data from two published studies: the
first aims to detect periodic genes of the organism Schizosaccharomyces
pombe, and the second aims to identify genes associated with prostate cancer.
\end{abstract}

%

\end{frontmatter}

\section{Introduction}
\label{sec:intro}

In current genetics and biological research, frequently thousands of
hypothesis tests are performed simultaneously. With such a large number
of tests, control of the family-wise error rate (FWER) is overly
conservative, resulting in low power for detection of true alternative
signals. For this reason, False Discovery Rate (FDR) control is an
intensely studied topic of interest. Methods such as the
Benjamini--Hochberg (B--H) procedure [\citet{benjamini95}] control FDR
when each hypothesis test is associated with a single $p$-value. Many
refinements have been proposed to increase power
[\citet{ghosh11,benjamini00,benjamini06,storey02}]. \citet{pounds06a} gives a good
summary of these approaches. However, when multiple $p$-values are
associated with each hypothesis, these methods alone are insufficient
for declaring significance while controlling FDR.

Multiple $p$-values can be considered for each hypothesis test in
settings such as repeated experiments in the same study, meta-analysis
across multiple studies and measurement of multiple aspects of a single
underlying process of interest. To perform a single hypothesis test
using multiple $p$-values, techniques for $p$-value combination are
frequently used to create a single summarized value. Multiple
comparison adjustments can then be applied to these summarized values.

With this move from single $p$-values to vectors of $p$-values, which we
refer to here as $p$-vectors, clear specification of the null and
alternative hypotheses is critical. If the goal is to pool information,
for example, when testing $p$-vectors from repeated experiments or from
multiple studies, then conjunction or partial conjunction hypotheses
are appropriate. The conjunction null hypothesis is that all $p$-values
contained in a $p$-vector are from a null distribution, and rejection is
possible when at least one $p$-value shows evidence of being from a
nonnull distribution. Rejection of the partial conjunction hypothesis
requires at least $u$ of $n$ $p$-values to show evidence of being from
nonnull distributions. There are scenarios, however, when the
hypothesis associated with each $p$-value of the $p$-vector is of interest
individually, and rejection should be possible only when there is
evidence that all such hypotheses are nonnull. In this case, the
disjunction hypothesis is of primary interest. Distinctions between the
conjunction, partial conjunction and disjunction hypotheses are further
described in Section~\ref{sec:disjunction} of this paper.

Testing of the disjunction hypothesis is appropriate when multiple
aspects of a single underlying biological process are measured. For
example, there is interest in detection of genes related to cancer
progression that are both differentially expressed on a transcriptomic
level and show evidence of copy number aberrations in cancerous tissue
[\citet{kim07,tsafrir06,fritz02,pollack02,tonon05}]. Another
motivating example is detection of periodic genes as explored by \citet
{lichtenberg05}. In this case, the disjunction hypothesis is considered
through one $p$-value for periodicity and a second for regulation of
expression. The most commonly used summary method for the disjunction
hypothesis uses the maximum of all $p$-values for each test [\citet
{wilkinson51}] and typically has very low power.

This paper presents an approach for $p$-value combination appropriate for
testing the disjunction hypothesis when there are two $p$-values
associated with each gene or location. The approach considers $p$-vectors
as locations on the unit square, where certain challenges absent in the
case of single $p$-values arise. First, the strict ordering of $p$-values
on the real line is lost. Second, relationships between $p$-vectors are
complicated and, third, their components may be correlated. In light of
these challenges, a method for large-scale simultaneous testing of the
disjunction hypothesis must accomplish three objectives. It must
account for the relative positioning of the $p$-vectors in the plane,
allow for multiple ordering schemes and, finally, allow for FDR control
under any correlation structure of the test statistics used to
calculate the $p$-vectors' components.

The approach proposed here addresses these challenges through the use
of Voronoi diagrams, flexible incorporation of ordering schemes and
empirical null distributions [\citet{efron04}]. This paper is organized
as follows. Section~\ref{sec:disjunction} details the disjunction
hypothesis framework while Section~\ref{sec:review} describes
background on control of FDR and existing univariate procedures.
Sections~\ref{sec:voronoi} and \ref{sec:orderingschemes} introduce
Voronoi diagrams and multiple ordering schemes for $p$-vectors.
Section~\ref{sec:sumandsig} describes a technique for summarizing
$p$-vectors to
a single value and details how these values can be used to control FDR.
We explore properties of the procedure through simulations in
Section~\ref{sec:depsims}. A~possible extension for higher dimensional
$p$-vectors is discussed in Section~\ref{sec:extension}. We apply the
procedure to two genomic studies in Sections~\ref{sec:applicationsoliva} and \ref{sec:applicationskim}.

\section{The disjunction of null hypotheses}
\label{sec:disjunction}

When multiple $p$-values are associated with a hypothesis test, the
interpretation of significance depends on the specification of null and
alternative hypotheses. Consider $m$ $p$-vectors, each of length~$n$, denoted
%
\begin{equation}
P_i=(p_{i1},\ldots,p_{in}), \qquad i=1,\ldots,m.
\end{equation}
In the context of a genomic study, $i$ is the index of the individual
gene, while $n$ is the number of $p$-values associated with each gene. We
employ notation used by \citet{benjamini08} to describe null and
alternative hypotheses. Testing the global null hypothesis, also known
as the conjunction of the null hypotheses, is equivalent to testing
that \emph{at least one} of the $p$-values $p_{i1},\ldots,p_{in}$ is significant:
\begin{eqnarray*}
&&H_0^{1/n}\!\dvtx\quad \mbox{all hypotheses associated with
$P_i$ are null},
\\
&&H_A^{1/n}\!\dvtx\quad \mbox{at least one hypothesis associated
with $P_i$ is nonnull.}
\end{eqnarray*}
$P$-value combination methods for testing the conjunction null include
the well-known Fisher's and Stouffer's methods for combining $p$-values
[\citet{fisher32,stouffer49}]. A comparison of these and other methods
is presented by \citet{loughin04}. Rejection of the conjunction null
can result from the influence of a single highly significant $p$-value
even when all other $p$-values show no evidence for the alternative
hypothesis. In this setting, the scientific conclusion from rejection
is not as strong it would be if a level of increased consistency across
$p$-values was enforced.

%
\begin{table}[b]
\caption{Outcomes for testing $m$ hypotheses}
\label{tab:testoutcomes}
\begin{tabular*}{\textwidth}{@{\extracolsep{\fill}}lccc@{}}
\hline
&\textbf{Fail to reject}&\textbf{Reject}&\multicolumn{1}{c@{}}{\textbf{Total}}\\
\hline
True null&$U$&$V$&$m_0$\\
True alternative&$T$&$S$&$m-m_0$\\[3pt]
Total&$m-R$&$R$&$m$\\
\hline
\end{tabular*}
\end{table}

\citeauthor{benjamini08} proposed techniques for addressing this
weakness through testing of the partial conjunction hypothesis:
\begin{eqnarray*}
&&H_0^{u/n}\!\dvtx\quad \mbox{at least $u$ of $n$ hypotheses
associated with $P_i$ are null},
\\
&&H_A^{u/n}\!\dvtx\quad \mbox{at least $u$ of $n$ hypotheses
associated with $P_i$ are nonnull.}
\end{eqnarray*}
This hypothesis requires a level of consistency of evidence across $u$
studies that is unnecessary in the conjunction framework, while still
allowing lack of significance for some associated $p$-values. It can be
considered a compromise between the conjunction and disjunction
hypotheses. The disjunction hypothesis is also referred to as the
disjunction of the null hypotheses and can be expressed as follows:
\begin{eqnarray*}
&&H_0^{n/n}\!\dvtx\quad\mbox{at least one hypothesis associated
with $P_i$ is null},
\\
&&H_A^{n/n}\!\dvtx\quad \mbox{all hypotheses associated with
$P_i$ are nonnull.}
\end{eqnarray*}
This hypothesis is desirable when considering multiple $p$-values per
test that are each of individual interest. The established $p$-value
combination approach for testing the disjunction hypothesis is to
simply select the maximum $p$-value of each $p$-vector [\citet
{wilkinson51}]. Error control procedures can then be applied to these
maximum values. This approach is generally conservative and exhibits
low power. The procedure described in this paper is suitable for
testing the disjunction hypothesis and results in a gain of power over
the maximum method of the $p$-value combination.

\section{The False Discovery Rate and review of existing procedures}
\label{sec:review}
To define FDR we consider Table~\ref{tab:testoutcomes}. The False
Discovery Rate is defined to be the \emph{expected proportion of false
rejections to total rejections}, or $E [V/R ]$ for $R>0$, and
0 if \mbox{$R=0$}. The FWER is defined to be the \emph{probability of at least
one false rejection}, $P(V>0)$. Particularly in studies testing
thousands or tens of thousands of hypotheses simultaneously, control of
FDR grants additional power for rejection relative to control of FWER.
Allowance of a controlled proportion of false positives enables
increased detection of more true signals relative to limiting
rejections based on the probability of at least one false rejection as
required by FWER control.

We next provide a brief review of two existing procedures that control
FDR when each hypothesis test has exactly one $p$-value: the
Benjamini--Hochberg (\mbox{B--H}) procedure and the Generalized
Benjamini--Hochberg procedure. The latter motivates our approach to
$p$-value combination for two-dimensional $p$-vectors.

\subsection{The Benjamini--Hochberg procedure and its generalization}
\label{subsec:bhprocedure}
Proposed by \citet{benjamini95}, the B--H procedure works as follows.
Assume $m$ continuous $p$-values $p_1,\ldots,p_m$ and that low values
indicate evidence against the null. Order them as $p_{(1)},\ldots
,p_{(m)}$ and compare to the thresholds $\alpha/m,2\alpha/m,\ldots
,\alpha
$. Define
%
\begin{equation}
\label{eq:BHdef}\hat{k}=\max \biggl\{i\dvtx p_{(i)}\le\frac{i}{m}
\alpha \biggr\}.
\end{equation}
If the set in (\ref{eq:BHdef}) is nonempty, reject the hypotheses
associated with $p_{(1)}$ through $p_{(\hat{k})}$, otherwise reject
nothing. \citeauthor{benjamini95} show that this procedure controls the
FDR at a nominal level $\alpha$.

\citet{ghosh11} proposed a family of testing procedures based on the
spacings of $p$-values. We present the basic procedure here. Again assume
$m$ independent $p$-values and order them as $p_{(1)},\ldots,p_{(m)}$.
Define the spacings [\citet{pyke65}] as
%
\begin{equation}
\label{eq:spacingsdef}\tilde{p}_i=p_{(i)}-p_{(i-1)},\qquad i=1,\ldots
,m+1,
\end{equation}
where $p_{(0)}=0$ and $p_{(m+1)}=1$. Under the null hypothesis the
original $p$-values are distributed as $\operatorname{Uniform}(0,1)$, whence the spacings
$\tilde{p}_i$ are marginally distributed as $\operatorname{Beta}(1,m)$. It is also
simple to calculate certain expectations. Specifically, $E[\tilde
{p}_i]=(m+1)^{-1}=E[\tilde{p}_1]$.

Recall that the B--H procedure defines $\hat{k}$ by comparing $p_{(1)},
p_{(2)}, \ldots, p_{(m)}$ to $\alpha/m,2\alpha/m,\ldots,\alpha$. These
quantities can be redefined as
%
\begin{equation}\quad
p_{(i)}=\sum_{j=1}^i
\tilde{p}_j\quad \mbox{and}\quad \frac
{i}{m}\alpha =i\cdot
\frac{m+1}{m}\cdot\frac{1}{m+1}\cdot\alpha=i\cdot\frac{m+1}{m} \cdot
E[\tilde{p}_1]\cdot\alpha.
\end{equation}
By substituting these quantities in the B--H procedure, the original
$\hat{k}$ can be rewritten in terms of the spacings as follows
[\citeauthor{ghosh11} (\citeyear{ghosh11,ghosh12})]:
%
\begin{equation}
\hat{k}=\max \Biggl\{i\dvtx\frac{1}{i}\sum_{j=1}^i
\tilde{p}_j\le\frac
{m+1}{m} E[\tilde{p}_1]\cdot
\alpha \Biggr\}. \label{eq:khatgen}
\end{equation}
According to \citeauthor{benjamini95}, this procedure controls FDR
at\break
$\alpha m_0/m$, where $m_0$ is the number of $p$-values associated with
the null hypothesis. In equation (\ref{eq:khatgen}) there is an extra
factor $(m+1)/m$. Elimination of this factor results in a slightly more
conservative procedure that preserves FDR control. Thus, $\tilde{k}$ is
defined as
%
\begin{equation}
\tilde{k}=\max \Biggl\{i\dvtx\frac{1}{i}\sum_{j=1}^i
\tilde{p}_j\le\alpha E[\tilde{p}_1] \Biggr\}.
\label{eq:khatgenbh}
\end{equation}
The definition of $\tilde{k}$ in (\ref{eq:khatgenbh}) hinges on a
comparison of the average spacings between ordered $p$-values to the
value $\alpha E[\tilde{p}_1]$. When there are numerous significant
$p$-values their spacings are small in comparison to the expectation of a
spacing under the null hypothesis. Detecting this change from the small
spacings of significant $p$-values to the larger spacings of null
$p$-values is also the motivation for the procedure described in this paper.

\section{The Voronoi diagram}
\label{sec:voronoi}
In the plane the spacings of two-dimensional $p$-vectors are more
difficult to characterize. One generalization of ``spacing'' is the
Voronoi diagram. It is a partition of the plane generated by an input
set of points that creates a cell around each input consisting of the
set of all points closer to that input than to any other. The basic
properties are described by \citet{okabe09}.

\begin{figure}[b]

\includegraphics{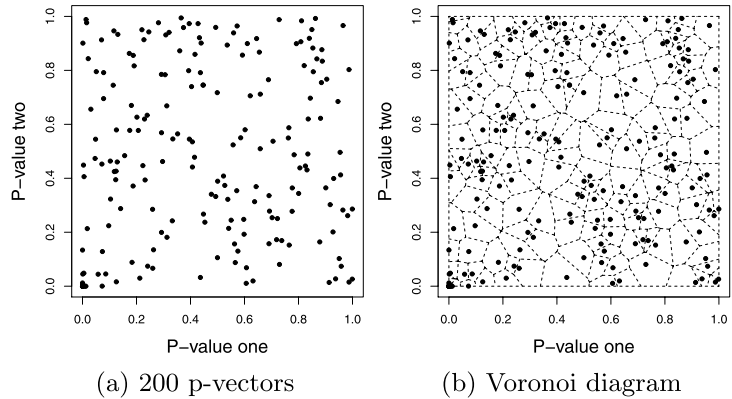}

\caption{\textup{(a)} 200 simulated $p$-vectors and \textup{(b)} the corresponding Voronoi
diagram.}
\label{fig:histograms}
\end{figure}

In the setting of two-dimensional $p$-vectors, the Voronoi diagram
partitions the unit square. For each $p$-vector, $P_i$, the diagram
creates a cell, $C_i$, consisting of all points closer to $P_i$ than to
any other $p$-vector. As \citet{jimenez02} discuss, Voronoi diagrams are
suitable for extension of the concept of one-dimensional spacings into
higher dimensions. An illustration of the diagram for a sample set of
points is presented in Figure~\ref{fig:histograms}(b). We follow this example
in Section~\ref{sec:orderingschemes} before switching to a larger
sample for subsequent sections.

%
Voronoi cells have many desirable properties that extend spacings to
the plane. Their area and shape reflect the relative positioning of the
input points. For example, clusters of inputs will have smaller cell
areas than uniformly distributed inputs. Similarly, if the inputs have
correlated components, there will be an increased concentration of
$p$-vectors along the diagonal of the unit square. The Voronoi cells for
the inputs along this diagonal will be smaller than those near the edge
of the clustering. Our procedure uses the areas of the Voronoi cells
generated by the set of $p$-vectors to account for their relative
positions. To compute each diagram and calculate the cell areas, we use
the R package deldir developed by \citet{deldir13} and available
through CRAN (\url{http://cran.r-project.org/web/packages/deldir/index.html}).

\section{Multiple ordering schemes}
\label{sec:orderingschemes}
Recall from Section~\ref{subsec:bhprocedure} that spacings were defined
as the difference between consecutive $p$-values. This definition is
dependent upon an ordering of the $p$-values that is unique on the real
line. However, this uniqueness is lost when $p$-vectors are considered as
bivariate locations in the unit square. We present and test multiple
ordering schemes for the plane, while continuing to assume that small
values of $p_{i1}$ and $p_{i2}$ indicate evidence against the null. For
this reason the orderings begin at the origin, and each $p$-vector is
ranked according to increasing values of $D$, its distance from the
origin. Each scheme defines $D$ differently. Thus, for each definition
$P_{(1)}$ is the $p$-vector with the smallest value of $D$, and $P_{(m)}$
with the largest. Here we describe the definition of $D$ for each scheme:
\begin{longlist}[1.]
\item[1.]\textit{Euclidean ordering} results in a movement from the origin
in contours with the shape of circles. Define $D^{(E)}$ as the
Euclidean distance from the origin
%
\begin{equation}
D_i^{(E)}=\sqrt{p_{i1}^2+p_{i2}^2}\qquad
(i=1,\ldots,m).
\end{equation}

\item[2.]\textit{Maximum ordering} results in contours with the shape of
squares. Define $D^{(M)}$ as
%
\begin{equation}
D_i^{(M)}=\max\{p_{i1},p_{i2}\}\qquad (i=1,
\ldots,m).
\end{equation}

\item[3.]\textit{Summation ordering} is equivalent to beginning at the
origin and moving out in contours of right isosceles triangles. In this
case, $D^{(S)}$ is defined as
%
\begin{equation}
D_i^{(S)}=p_{i1}+p_{i2}\qquad (i=1,\ldots,m).
\end{equation}

\item[4.]\textit{de Lichtenberg ordering} is a ranking scheme proposed by
\citet{lichtenberg05}. The scheme defines
%
\begin{equation}\qquad
D^{(L)}_i=p_{i1} p_{i2} \biggl(1+
\biggl(\frac
{p_{i1}}{0.001} \biggr)^2 \biggr) \biggl(1+ \biggl(
\frac{p_{i2}}{0.001} \biggr)^2 \biggr)\qquad (i=1,\ldots,m).
\end{equation}
Note that $D^{(L)}$ consists of four multiplicative factors. The first
two weight $D^{(L)}$ according to the value of each individual
component, and the last two penalize $p$-vectors that have only one very
small component. For typical $p$-vectors the values for $D^{(L)}$ are
very large as a result of division by 0.001 of both $p_{i1}$ and
$p_{i2}$. This magnitude is not a concern, as the interest is only in
their relative values for the purpose of ranking and the values
themselves are not of particular interest. The contour lines for this
ordering scheme move from the origin in lines approximating an inverse
function such as $y=1/(x^3)$.
\end{longlist}

\begin{figure}

\includegraphics{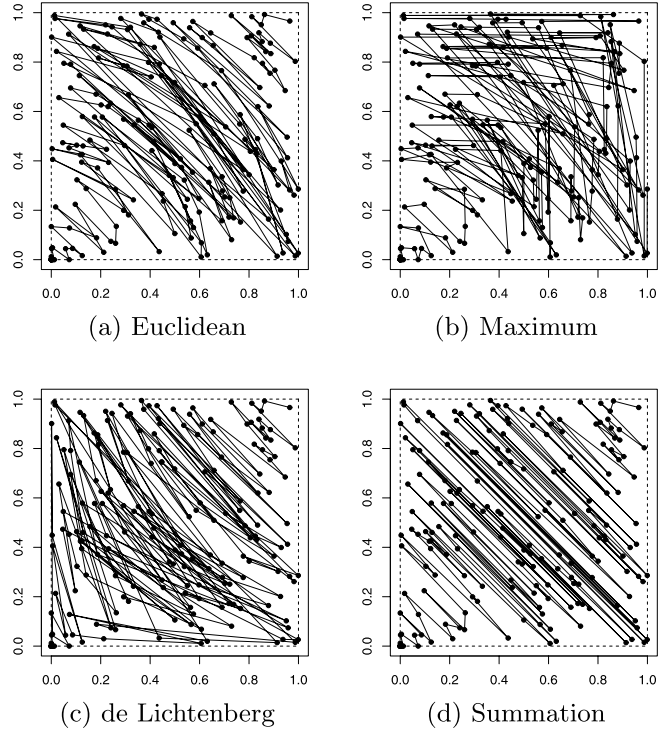}

\caption{Illustration of all four ordering schemes with the sample set
of 200 $p$-vectors. The solid lines join $p$-vectors that are consecutively
ranked, that is, $P_{(1)}$ to $P_{(2)}$, $P_{(2)}$ to $P_{(3)}$, etc.}
\label{fig:orderings}
\end{figure}

Figure~\ref{fig:orderings} illustrates these four ordering schemes
using the sample set of 200 $p$-vectors from Section~\ref{sec:voronoi}.
Table~\ref{tab:orderingexample} presents a numerical example using five
$p$-vectors.

%

\begin{table}
\caption{Example of ordering results for five $p$-vectors. For each
ranking scheme the distance, $D$, is presented for each $p$-vector along
with the resulting rank in parentheses}
\label{tab:orderingexample}
\begin{tabular*}{\textwidth}{@{\extracolsep{\fill}}lcccccccc@{}}
\hline
$\bolds{P_i}$ & $\bolds{D_i^{(E)}}$& & $\bolds{D_i^{(M)}}$ & &
$\bolds{D_i^{(S)}}$ & & $\bolds{D_i^{(L)}}$ & \\
\hline
$(0.85, 0.51)$ & 0.99 & (3) & 0.85 & (3) & 1.36 & (4) & $8.1\cdot
10^{10}$ & (4) \\
$(0.91, 0.80)$ & 1.21 & (5) & 0.91 & (4) & 1.71 & (5) & $3.9\cdot
10^{11}$ & (5) \\
$(0.23, 0.97)$ & 1.00 & (4) & 0.97 & (5) & 1.20 & (3) & $1.1\cdot
10^{10}$ & (3) \\
$(0.62, 0.34)$ & 0.71 & (2) & 0.62 & (1) & 0.96 & (2) & $9.5\cdot10^9$
& (2) \\
$(0.07, 0.63)$ & 0.63 & (1) & 0.63 & (2) & 0.79 & (1) & $8.9\cdot10^7$
& (1)\\
\hline
\end{tabular*}
\end{table}

\begin{figure}[b]

\includegraphics{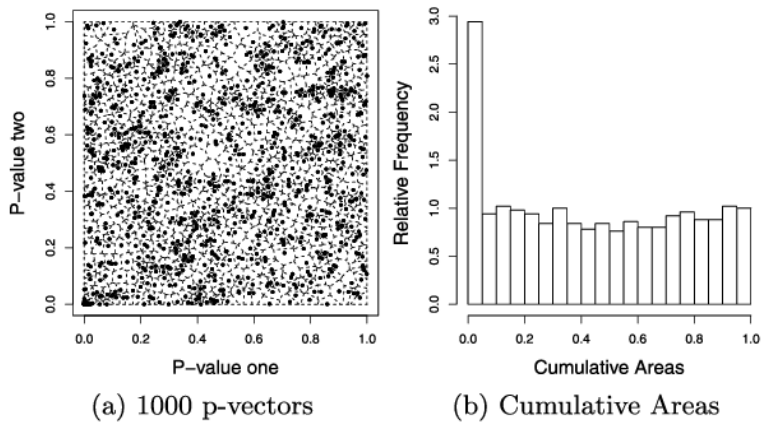}

\caption{An example of \textup{(a)} 1000 simulated $p$-vectors with independent
components and \textup{(b)} a histogram of their cumulative areas calculated
using the Euclidean ordering scheme. The sharp spike in the histogram
corresponds to the $p$-vectors associated with alternative hypotheses.}
\label{fig:indexample}
\end{figure}

It is noteworthy that three of the four ranking schemes described have
concave contour lines: Euclidean, Maximum and Summation. The remaining
scheme, de Lichtenberg, has convex contour lines. As we will see, these
characteristics have important implications for error control.

\section{Summarizing $p$-vectors and declaring significance}
\label{sec:sumandsig}

The described ranking schemes can be combined with Voronoi cell areas
to summarize each ranked $p$-vector as a single value in the interval
$(0,1)$. Define $A_{(i)}$ as the area of the Voronoi cell associated with
the ordered $p$-vector $P_{(i)}$ and the cumulative sum of these ordered
areas as
%
\begin{equation}
\label{eq:Ti} T_{(i)}=\sum_{j=1}^i
A_{(j)} \qquad(i=1,\ldots,m).
\end{equation}
These cumulative sums serve as combined $p$-values in an analogous manner
that cumulative spacings comprise $p$-values in one dimension. These
$T_{(i)}$ reflect both the relative positioning of the $p$-vectors in
space and their distance from the origin. They can be used to make
decisions in the hypothesis testing framework. Figure~\ref{fig:indexample} illustrates a sample set of 1000 $p$-vectors and a
histogram of their cumulative areas. In this example the components of
the $p$-vectors are independent, and 10\% of $p$-vectors are associated
with an alternative hypothesis.

%
\begin{table}
\caption{Power (1-NDR) results of simulation studies under independence}
\label{tab:indndr}
\begin{tabular*}{\textwidth}{@{\extracolsep{\fill}}lccccc@{}}
\hline
& \textbf{Euclidean} & \textbf{Maximum} & \textbf{Summation} &
\textbf{de Lichtenberg} & \multicolumn{1}{c@{}}{\textbf{Existing procedure}} \\
\hline
$\mu_{A}=2$ & 0.200 & 0.189 & 0.216 & 0.205 & 0.005 \\
$\mu_{A}=3$ & 0.772 & 0.760 & 0.788 & 0.796 & 0.098 \\
$\mu_{A}=4$ & 0.976 & 0.975 & 0.977 & 0.979 & 0.744 \\
\hline
\end{tabular*}
\end{table}

%

\subsection{Multiple hypothesis testing under independence}
\label{subsec:independence}
When the components of the $p$-vectors are assumed to be independent,
standard multiple comparisons procedures such as B--H can be applied to
the cumulative spacings with very good results. Simulation studies were
performed to test the properties of FDR control and the power of this
method. For each study 100 sets of test statistics were generated
according to
\[
(t_{i1},t_{i2})\sim \operatorname{MVN} \left( %
\pmatrix{
\mu_i\vspace*{2pt}
\cr
\mu_i} %
, %
\pmatrix{1&0\vspace*{2pt}
\cr
0&1 } %
\right),\qquad i=1,\ldots,2000;
\]
10\% of statistics were associated with an alternative hypothesis ($\mu
_i=\mu_{A}$), while the remaining 90\% were null ($\mu_i=0$). $P$-vectors
were formed from 2-sided $p$-values according to
\[
P_i=(p_{i1},p_{i2})= \bigl(P\bigl(Z>|t_{i1}|\bigr),P\bigl(Z>|t_{i2}|\bigr)
\bigr)\qquad \mbox{where } Z\sim N(0,1)
\]
for $i=1,\ldots,2000$. The proposed method was applied to each data set,
using the four described ordering schemes from Section~\ref{sec:orderingschemes}. Additionally, the existing $p$-value combination
technique based on the maximum was applied. After applying the B--H
procedure to each set of summary values, the FDR and 1-non discovery
rate (NDR) was recorded. Using notation from Table~\ref{tab:testoutcomes}, 1-NDR is defined as $E [S/(S+T) ]$. This
quantity can be viewed as a measure of power.

Tables~\ref{tab:indndr} and \ref{tab:indfdr} summarize the results for
studies where $\mu_A=2, 3, 4$, respectively. The results show that,
under all ordering schemes, the proposed combination method results in
greatly increased power. All concave schemes (Euclidean, Maximum and
Summation) control FDR at the desired level $\alpha=0.05$, but the
convex de Lichtenberg scheme does not. This difference becomes more
pronounced when $p$-vectors with correlated components are considered.
Additional simulations using correlated test statistics show that
application of the B--H procedure to combined values is insufficient to
control FDR when the correlation between test statistics surpasses 0.2.
This loss of FDR control is a result of the increased concentration of
$p$-vectors along the diagonal of the unit square under correlation,
which changes the characteristics of the cumulative areas. In
Section~\ref{subsec:dependence} we discuss approaches appropriate for multiple
testing in these conditions.

\begin{table}
\caption{False Discovery Rate results of simulation studies under independence}
\label{tab:indfdr}
\begin{tabular*}{\textwidth}{@{\extracolsep{\fill}}lccccc@{}}
\hline
& \textbf{Euclidean} & \textbf{Maximum} & \textbf{Summation} &
\textbf{de Lichtenberg} & \multicolumn{1}{c@{}}{\textbf{Existing procedure}} \\
\hline
$\mu_{A}=2$& 0.041 & 0.037 & 0.048 & 0.041 & 0.000 \\
$\mu_{A}=3$ & 0.042 & 0.038 & 0.049 & 0.056 & 0.000 \\
$\mu_{A}=4$ & 0.042 & 0.040 & 0.045 & 0.053 & 0.000 \\
\hline
\end{tabular*}
\end{table}

%
\subsection{Multiple hypothesis testing under dependence}
\label{subsec:dependence}

In certain settings the individual components of $p$-vectors may be
correlated. For example, correlation between components may occur when
different but related aspects of an underlying biological process are
measured. Any technique used for testing the disjunction hypothesis in
this setting should be robust to this structure. Using an empirical
null approach [\citeauthor{efron04} (\citeyear{efron04,efron07})] in the place of the B--H
procedure results in FDR control for all positive correlation
structures, although the trade-off is decreased power in the case of
independent components.

The use of an empirical null for determining statistical significance
was proposed by \citet{efron04}. We consider a transformation of the
summarized cumulative areas $T_{(i)}$ as defined in (\ref{eq:Ti}):
%
\begin{equation}
\label{eq:Zi} Z_{(i)}=\Phi^{-1}(T_{(i)}) \qquad(i=1,
\ldots,m),
\end{equation}
where $\Phi$ is the cumulative distribution function for the standard
normal random variable. In the empirical null framework, these
transformed values are assumed to be from a mixture distribution
%
\begin{equation}
\label{eq:Zidist} Z_{(i)}\sim f(z) = \delta f_0(z) + (1-\delta
)f_1(z)\qquad (i=1,\ldots,m),
\end{equation}
where $f_0$ is the null, or ``uninteresting'' distribution, and $f_1$
is the alternative, or ``interesting'' distribution. Under the
theoretical null hypothesis, $f_0$ is the $N(0,1)$ distribution, however,
in large-scale multiple testing problems the majority of values may
behave differently. When this is the case, use of an empirically
determined $f_0$ in place of $N(0,1)$ has important implications for
the resulting inference. For example, if $f_0$ is estimated to be
$N(0,1.3)$, then inference based on the assumptions of $N(0,1)$ would
result in elevated type I error. Similarly, if the empirical null has
smaller variance, then its use results in a gain of power without
sacrificing type I error control.

\begin{figure}

\includegraphics{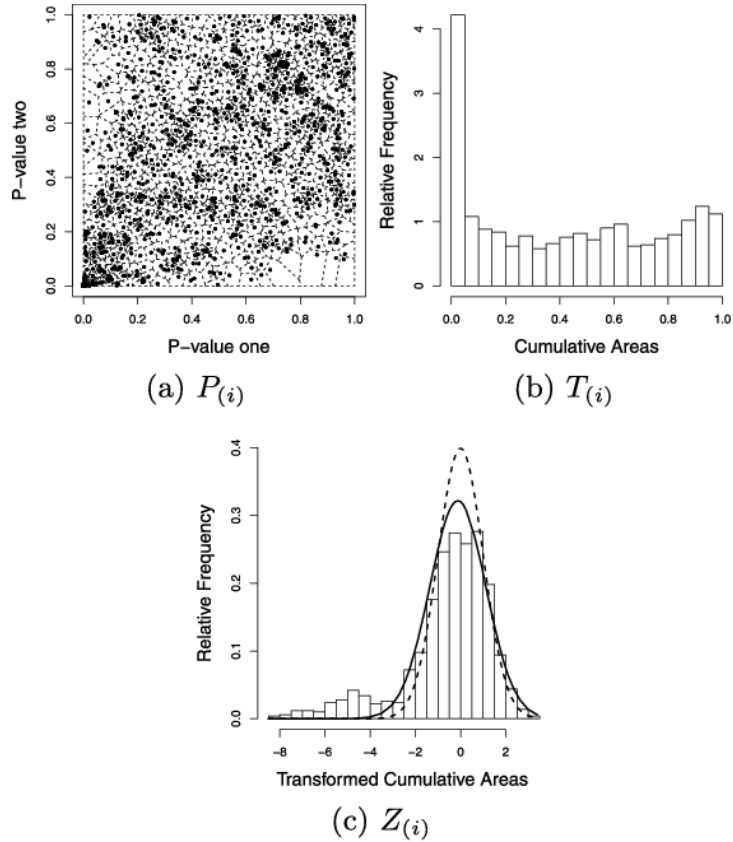}

\caption{An example of \textup{(a)} 1000 simulated $p$-vectors with dependent
components, \textup{(b)} a histogram of their cumulative areas calculated using
the Euclidean ordering scheme, and \textup{(c)} a histogram of the transformed
cumulative areas with empirical (solid line) and theoretical (dashed
line) null distributions.}
\label{fig:depexample} 
\end{figure}

%
Figure~\ref{fig:depexample} illustrates the effect of the
transformation from $T_{(i)}$ to $Z_{(i)}$ for a sample set of 1000
$p$-vectors. The transformation makes it much easier to detect deviations
from the null hypothesis, as true alternative hypotheses are presented
as a second, smaller peak to the left of the null distribution instead
of in a single spike for the original values. The bivariate test
statistics used to calculate these $p$-vectors had a correlation of 0.7.
The histogram of transformed values in Figure~\ref{fig:depexample}(c) shows
evidence of a null distribution that differs from $N(0,1)$, as the
dependence structure of the $p$-vectors results in thicker tails than the
theoretical null predicts. For this reason, it is desirable to use an
empirical null as a basis for our inference when the components of the
$p$-vectors show evidence of correlation.

Given the choice of an empirical null, there are several options for
declaring significance while controlling FDR. \citet{efron04} defined
the \emph{local false discovery rate} (fdr) as the posterior
probability of a value $z$ being from the null distribution, given the
value of $z$:
%
\begin{equation}
\operatorname{fdr}(z)=P (z_i\sim f_0|z_i=z ).
\end{equation}
Another important probability that can be estimated using an empirical
null is the left-tail False Discovery Rate. For each value $z$ the
corresponding left-tail FDR is defined as
%
\begin{equation}
\operatorname{FDR}(z) = P (z_i\sim f_0|z_i\le z ).
\end{equation}
Inference can be made based on estimated fdr or FDR values. A variety
of approaches have been developed for estimation of empirical null
distributions and related values [\citet
{muralidharan10,efron04,pounds03,strimmer08,jin07}]. We use an R
package (mixFdr) developed by \citet{muralidharan10} and available
through CRAN (\url{http://cran.r-project.org/web/packages/mixfdr}),
which uses an empirical Bayes mixture method to fit an empirical null,
estimate effect sizes, fdr and FDR. The use of other packages or
techniques is certainly possible. The function we used in simulation
studies, mixFdr, includes two tuning parameters: $J$, the number of
distributions to be estimated, and $P$, a penalization parameter. A
higher value of $P$ encourages estimation of a larger null group and
closer estimation of the central peak.

Careful calibration of $J$ and $P$, and even experimentation with other
techniques for empirical null estimation, are desirable when a single
data set is under consideration. The function mixFdr estimates the
left-tail False Discovery Rate for each $Z_{(i)}$, and we declare
significant all $p$-vectors with these estimates of left-tail FDR less
than 0.05. This approach results in appropriate error control that is
robust to correlation in the components of $p$-vectors. Section~\ref{sec:depsims} describes a simulation study performed to explore
properties of power and FDR control when this approach is applied to
the cumulative areas from ordered $p$-vectors.

\section{Simulation study when components are correlated}
\label{sec:depsims}
To illustrate properties of the procedure, we ran three simulation
studies: one each for strong, moderate and weak alternative signals. We
were interested in evaluating FDR control and power. For each simulated
data set we set $m=2000$ and generated test statistics by
\[
(t_{i1},t_{i2})\sim \operatorname{MVN}\biggl ( %
\pmatrix{
\mu_i\vspace*{2pt}
\cr
\mu_i } %
, %
\pmatrix{1&\rho\vspace*{2pt}
\cr
\rho&1 } %
\biggr),\qquad i=1,\ldots,m
\]
for strong, moderate and weak alternative signals $\mu_A=4, 3, 2$,
respectively. For null test statistics, $\mu_i=0$. 10\% of test
statistics for each data set were generated from the alternative
distribution. $P$-vectors were formed from 2-sided $p$-values.

\begin{figure}

\includegraphics{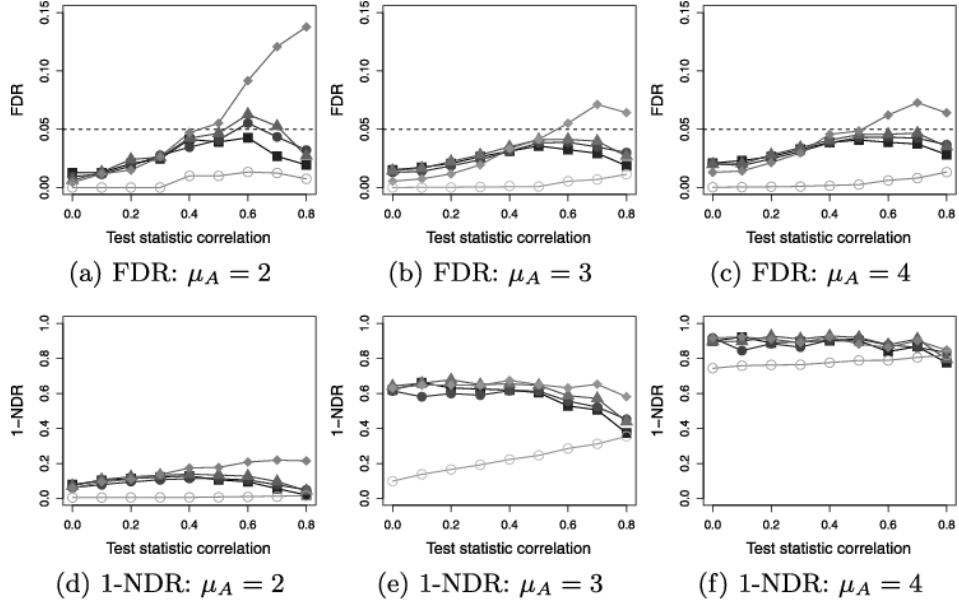}

\caption{Summarized results of simulation studies for test statistics
with varying correlation structure. \textup{(a)}, \textup{(b)} and \textup{(c)} present FDR when
alternative signals are 2, 3 and 4 standard deviations from the null.
\textup{(d)}, \textup{(e)} and \textup{(f)} present 1-NDR for the same data sets. Solid squares,
circles, triangles and diamonds represent Euclidean, Maximum, Summation
and de Lichtenberg ordering schemes, respectively. The open circles
represent the existing approach using the maximum of each component for
inference.}
\label{fig:depsims}
\end{figure}

For each simulation study $\rho$ varied from 0 to 0.8 in increments of
0.1, and 100 data sets were simulated for each value of $\rho$. We
performed the procedure using all four ordering schemes on each
simulated data set, using mixFdr to estimate empirical null
distributions and left-tail FDR for each data set, rejecting all
hypotheses associated with $p$-vectors whose estimated left-tail FDR was
less than 0.05. We set $J=2$ for all data sets, and after calibrating
the fit of several example empirical nulls for weak, moderate and
strong signals, we set $P=400$, 800 and 1000 for the respective
simulations. Additionally, the B--H procedure was applied to the maximum
values from each $p$-vector to compare the proposed method to an existing
approach. Figure~\ref{fig:depsims} summarizes the results of all three
simulation studies. We include the R code used to perform these
simulations as a supplementary file [\citet{phillipssup2}].

%

The proposed technique for combining $p$-values has improved power when
compared to the existing procedure. This improvement is greatest for
weak and moderate alternatives. These simulations further show that
ordering scheme matters, although the differences between the three
convex (Euclidean, Maximum, Summation) ordering schemes are small
compared to the difference between them and the de Lichtenberg
ordering. The de Lichtenberg ordering displays characteristics that
differ considerably from the other three. Specifically, it shows a
tendency to lose FDR control as the components of the $p$-vectors become
more correlated. We present simulation results for this ordering for
the sake of completeness, but we do not recommend using it for data
analysis when testing the disjunction null. The nature of its contour
lines suggests this ordering scheme is in fact more appropriate for
testing the conjunction or partial conjunction hypothesis, as these
lines resemble the contour lines for Fisher's or Stouffer's $p$-value
combination techniques from Figure~1 of \citet{owen09}.

Further simulations using the empirical null allowed evaluation of the
proposed approach in the presence of $p$-vectors constructed from test
statistics with means ($0,\mu_A$) or $(\mu_A,0$). These $p$-vectors
should not be found significant in the disjunction setting, but should
be under the conjunction framework. In the presence of up to 20\% of
such vectors, and for correlation structures from $\rho=0$ to $\rho
=0.8$, FDR control was maintained below the nominal 0.05 level, and
power properties showed a clear advantage over existing methods.
Detailed descriptions and results are included in supplementary
materials [\citet{phillipssup1}].

\section{Extension to higher dimensions}
\label{sec:extension}

The approach described in this paper is suitable when there are two
$p$-values associated with each hypothesis test, however, in many
situations three or more $p$-values will be available. In theory, the
procedure can be extended to higher dimensions by replacing cumulative
areas with cumulative volumes, hypervolumes, etc. In practice, however,
the computation complexity for Voronoi cells increases quickly with
dimension. Average time complexity is as low as $\mathcal{O}(n)$ in the
plane, but is at least $\mathcal{O}(n^2)$ in 3-space [\citet{okabe09}].
To avoid this disadvantage, we consider an alternative extension using
the sets of all possible pairs of components. Consider a set of $m$
3-dimensional $p$-vectors:
%
\begin{equation}
P_i= (p_{i1},p_{i2},p_{i3} ),\qquad i=1,
\ldots,m.
\end{equation}
Then define three sets of 2-dimensional $p$-vectors constructed via a
pairwise combination of components of $P_i$:
%
\begin{equation}
\bigl\{ (p_{i1},p_{i2} ) \bigr\}, \bigl\{
(p_{i1},p_{i3} ) \bigr\}, \bigl\{ (p_{i2},p_{i3}
) \bigr\},\qquad i=1,\ldots,m.
\end{equation}
For each of these sets of two-dimensional $p$-vectors the Voronoi diagram
is computed and cell areas saved. Thus, each $p$-vector $P_i$ is
associated with three individual cell areas, $A_i^{1,2}, A_i^{1,3}$
and $A_i^{2,3}$, as well as an average area $\bar
{A}_i=(A_i^{1,2}+A_i^{1,3}+A_i^{2,3})/3$. This average area can then be
used in conjunction with an ordering scheme to create the summarized
areas used for inference. Define $P_{(1)}, P_{(2)},\ldots,P_{(m)}$ to
be the $p$-vectors ranked according to a specified ordering scheme
such\vadjust{\goodbreak}
as Euclidean distance from the origin, and $\bar{A}_{(1)}, \bar
{A}_{(2)},\ldots,\bar{A}_{(m)}$ to be the corresponding average areas.
Then the cumulative average areas are defined as $\bar{T}_{(i)}=\sum_{j=1}^i \bar{A}_{(i)}$.

Multiple testing can then be performed on these summarized cumulative
average areas using the methods described in Sections~\ref{subsec:independence} and \ref{subsec:dependence}. Further
investigation into the properties of this approach is necessary, as
well as research on other possible extensions for higher dimensions. A
preliminary simulation study using three-dimensional $p$-vectors with
independent components was conducted with weak, moderate and strong
alternative test statistics. For each data set, test statistics were
generated according to
\[
(t_{i1},t_{i2},t_{i3})\sim \operatorname{MVN} \left( %
\pmatrix{\mu_i\vspace*{2pt}
\cr
\mu_i\vspace*{2pt}\cr \mu_i }
, %
\pmatrix{1&0&0\vspace*{2pt}
\cr
0&1&0\vspace*{2pt}
\cr
0&0&1 } %
\right),\qquad i=1,\ldots,m.
\]
Three-dimensional $P$-vectors were formed from 2-sided $p$-values. The
resulting $p$-vectors were ordered according to Euclidean distance from
the origin. Hypothesis testing was performed using the B--H procedure on
the summarized cumulative average areas. The existing technique of
applying the B--H procedure to the set of maximum $p$-values from each
$p$-vector was also performed for comparison. Table~\ref{tab:3dsim}
summarizes the findings of the simulations.

\begin{table}
\caption{Simulation results for proposed extension}
\label{tab:3dsim}
\begin{tabular*}{\textwidth}{@{\extracolsep{\fill}}lcccccc@{}}
\hline
& \multicolumn{3}{c}{\textbf{Proposed extension}}& \multicolumn{3}{c@{}}{\textbf{Existing
approach}}\\[-6pt]
& \multicolumn{3}{c}{\hrulefill}& \multicolumn{3}{c@{}}{\hrulefill}\\
& $\bolds{\mu_A=2}$ &$\bolds{\mu_A=3}$ &$\bolds{\mu_A=4}$ &
$\bolds{\mu_A=2}$ &$\bolds{\mu_A=3}$ &\multicolumn{1}{c@{}}{$\bolds{\mu
_A=4}$ }\\
\hline
FDR & 0.023 & 0.004& 0.023 &0.000 & 0.000& 0.000\\
1-NDR& 0.098 &0.730 &0.986 & 0.005 & 0.007 & 0.610\\
\hline
\end{tabular*}
\end{table}

\section{Application to Schizosaccaromyces pombe data}
\label{sec:applicationsoliva}
In 2004 and 2005, three papers were published investigating the
periodicity of genes in the fission yeast cell \textit
{Schizosaccharomyces pombe}. Specifically, \citet{oliva05} produced
three data sets including time points for three complete cell cycles
using two different synchronization techniques. In their paper they
identified 750 genes determined to be periodically expressed based on a
ranking scheme and cutoff. We apply our approach to test for
periodicity in a hypothesis testing framework using Fisher's exact G
statistic to measure evidence of periodicity.

\subsection{The data}
\label{subsec:olivadata}
Three microarray data sets from \textit{Schizosaccharomyces pombe} from
\citeauthor{oliva05} were used: Elutriation a, Elutriation b and Cdc25.
The first two were produced using Elutriation synchronization, and the
last using a Cdc25 block-release synchronization technique. We apply
our technique on the two Elutriation sets, using Fisher's exact G
statistic to calculate the $p$-vector for each gene. This test statistic
requires evenly spaced time points, necessitating omission of any
measurements that occur at uneven intervals. The Elutriation a data set
includes 50 time points, however, only 33 are at regular intervals of 8
minutes. For Elutriation b, many of the time points are technical
repeats. We keep the first measurement in each case, leaving 31 evenly
spaced time points for each gene taken at intervals of 10 minutes. The
Cdc25 data set has a total of 51 evenly spaced time points, taken at
intervals of 15 minutes. Only genes with complete measurements for all
selected time points are considered.

\begin{table}
\caption{Summary of results from Oliva et~al. data sets
considered separately}\label{tab:olivasummary}
\begin{tabular*}{\textwidth}{@{\extracolsep{\fill}}lccc@{}}
\hline
&\textbf{Elutriation a} & \textbf{Elutriation b} & \multicolumn{1}{c@{}}{\textbf{Cdc25}} \\
\hline
Complete genes & 3050 & 2394 & 3724 \\
Evenly spaced time points & 33 & 31 & 51 \\
Genes (\%) with $p$-values $<$0.05, & 868 (22.8\%) & 546 (26.4\%) & 2458
(66.0\%)\\
Significant genes (\%) using B--H & 527 (17.3\%) & 155 (6.5\%) & 2252
(60.5\%)\\
\hline
\end{tabular*}
\end{table}

\begin{figure}[b]

\includegraphics{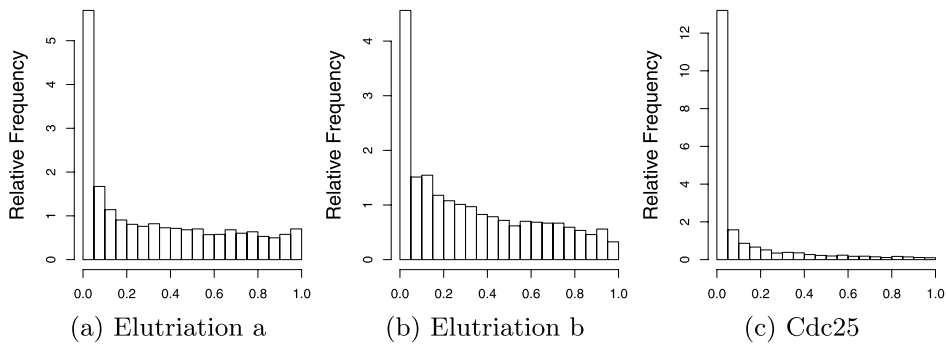}

\caption{Histogram of $p$-values for \textup{(a)} Elutriation~a, \textup{(b)}
Elutriation~b and \textup{(c)} Cdc25 block release. Note the strong evidence of periodicity
in all three experiments, particularly Cdc25.}
\label{fig:olivahists}
\end{figure}

\begin{figure}

\includegraphics{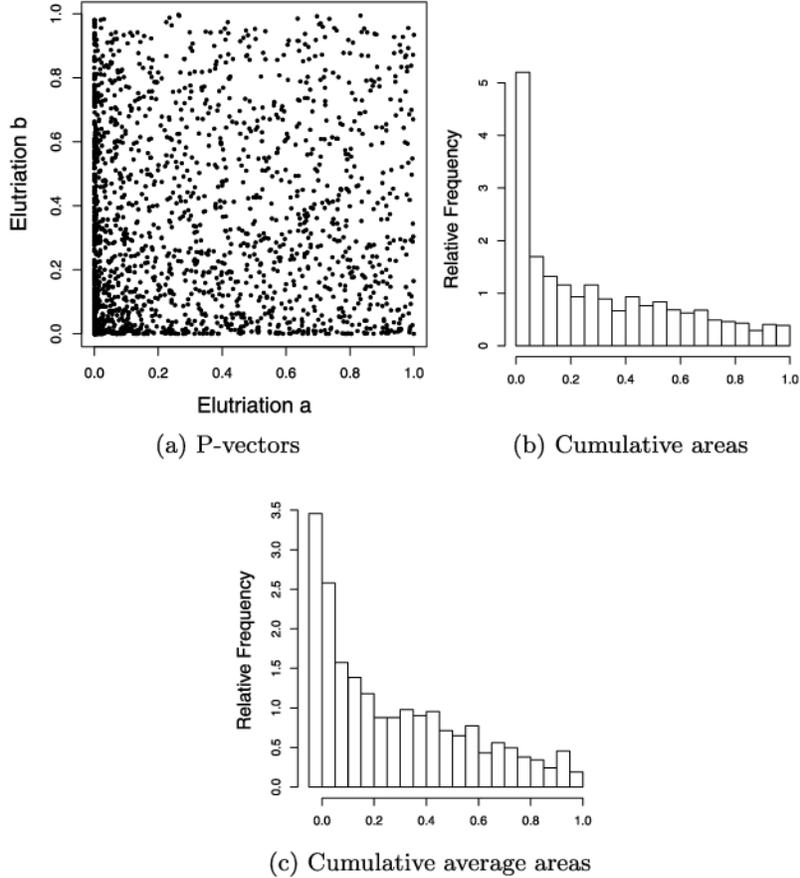}

\caption{\textup{(a)} $P$-vectors formed from Fisher's G statistic of
Elutriation~a and Elutriation~b, \textup{(b)}~a~histogram of cumulative cell areas formed
using the Euclidean ordering scheme, and \textup{(c)} a histogram of cumulative
average cell areas when all three experiments are considered.}
\label{fig:abEuclid}
\end{figure}

\subsection{Results using existing procedures}
\label{subsec:olivaexisting}
The data show evidence of wide\-spread periodicity. Considered
separately, Elutriation a, Elutriation b and Cdc25 have 22.8\%, 26.4\%
and 66\% of $p$-values less than 0.05. Even controlling FDR using the B--H
procedure on each set independently results in a very high rate of
rejection. Table~\ref{tab:olivasummary} presents a summary of the data
and marginal analysis of all three data sets. Figure~\ref{fig:olivahists} presents histograms of the $p$-values when the data sets
are considered independently.

%
%
Consider the $p$-vectors formed using $p$-values generated by Elutriation a
data and Elutriation b data. Note that these two Elutriation data sets
were both generated using the same synchronization technique, and the
$p$-values generated by each repetition have roughly comparable marginal
distributions.
To test the disjunction hypothesis for Elutriation a and Elutriation b,
using an existing technique the maximum $p$-value for each gene is
preserved. The B--H procedure is then applied to these maximum values.
The resulting number of rejections is 15, which is surprisingly low.
Figure~\ref{fig:abEuclid}(a) helps to explain this result. The
$p$-vectors' components do not show evidence of correlation, thus
considering only the maximum of each $p$-vector's components gives a
distribution that is very different from either of the marginal
distributions. Our proposed approach uses information from both
$p$-values and gives a different result.

\subsection{Results using Voronoi $p$-value combination on Elutriation data}
\label{subsec:elutab}
We apply our $p$-value combination method using the Euclidean, Maximum
and Summation ordering schemes to the $p$-vectors formed from the
Elutriation a and b experiments. The $p$-vectors are plotted in
Figure~\ref{fig:abEuclid}(a). The components of the $p$-vectors do
not show
evidence of high correlation, and we apply the B--H procedure to the
cumulative areas generated from each ordering scheme. This application
results in 225, 213 and 249 rejections of the disjunction hypothesis
using Euclidean, Maximum and Summation orderings, respectively.

%
%

Application of an empirical null approach to the combined areas yields
a very different result. Because of the high amount of periodicity
detected in the experiments, the empirical null is estimated to have a
negative mean. This shift to the left of up to $-0.87$ results in
rejection of far fewer genes: 15, 12 and 11 for the three concave
ordering schemes. These genes could be considered significantly more
periodic than the rest, although other genes also show evidence of periodicity.

The two considerations of the combined values reflect two different
scientific questions. By using the B--H procedure on the combined areas,
the genes found are those that show significant periodic expression in
both elutriation experiments. The genes found using the empirical null
procedure are those genes that are significantly periodic in both
experiments relative to the majority of genes. This distinction
explains the difference in numbers of genes found significant.

\subsection{Extension of procedure to include Cdc25}
\label{subsec:olivaextension}
The extension to three dimensions described in Section~\ref{sec:extension} can be applied to the 3-dimensional $p$-vectors formed
from \citeauthor{oliva05} data. We order the three-dimensional
$p$-vectors according to their Euclidean distance from the origin and
calculate the three Voronoi cell areas associated with each $p$-vector.
From these we calculate each $p$-vector's average cell area and then
cumulative average areas. Figure~\ref{fig:abEuclid}(c) presents a
histogram of these values. Note that many of these cumulative areas are
quite small as a result of the high number of very small $p$-values from
the Cdc25 experiment.

Application of the B--H procedure to the cumulative average areas formed
using the Euclidean ordering scheme results in rejection of 165
disjunction hypotheses. These 165 genes are those that show significant
evidence of periodic expression in all three of the experiments
performed by \citeauthor{oliva05} The existing procedure using the
maximum values yields a mere 12 rejections for these experiments. Using
an empirical null approach on the transformed cumulative average areas
yields results similar to those discussed in Section~\ref{subsec:elutab}. Because of the evidence of widespread periodicity
throughout the experiment, only 8 genes show behavior that is
significantly more periodic in comparison to the majority of genes when
all three experiments are considered.

\section{An application related to prostate cancer}
\label{sec:applicationskim}
Identification of genes implicated in cancer progression is a research
topic of great interest. Several studies have shown interest in
identifying genes that show both alterations in copy number and
evidence of differential expression in cancerous tumors
[\citet{kim07,tsafrir06,fritz02,pollack02,tonon05}]. We applied our method to data
produced by \citeauthor{kim07} in a study on prostate cancer
progression. Data on copy number and gene expression was gathered for
7534 genes using prostate cell populations from low-grade and
high-grade samples of cancerous tissue. For details on data acquisition
and cleaning see the \citeauthor{kim07} paper.

\begin{figure}

\includegraphics{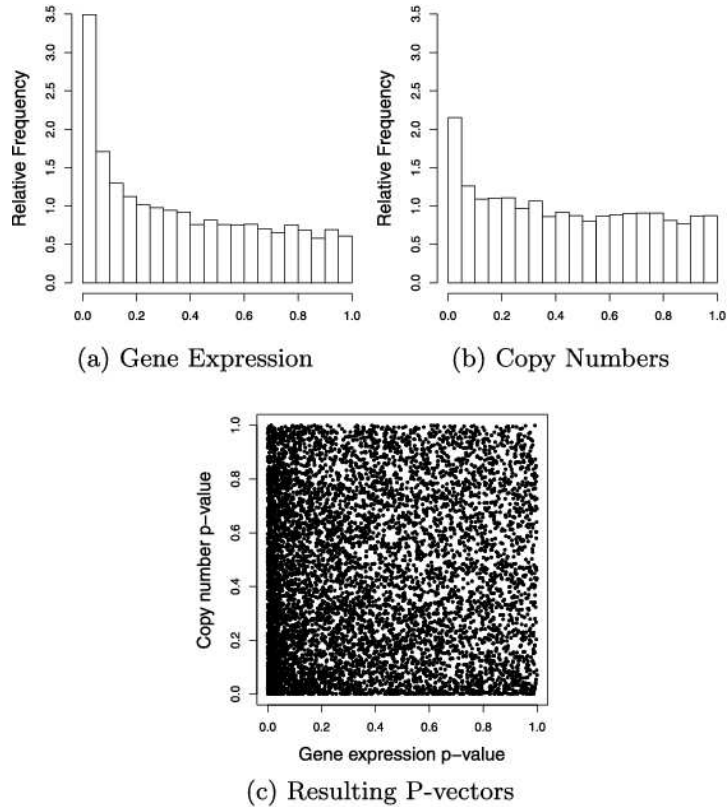}

\caption{Histogram of $p$-values for \textup{(a)} expression and \textup{(b)} copy number.
\textup{(c)} presents the resulting $p$-vectors in the unit square.}
\label{fig:kimdata}
\end{figure}

We calculated $t$-statistics for genetic expression and copy number
aberrations comparing tissue types. For each of 7534 genes we compute a
two-dimensional $p$-vector from the resulting 2-sided $p$-values based on
the $t$-statistics. Figures~\ref{fig:kimdata}(a) and \ref{fig:kimdata}(b)
present histograms of the expression and copy number $p$-values, while
Figure~\ref{fig:kimdata}(c) presents a representation of the
resulting $p$-vectors. Upon close inspection, it is revealed that the
smallest copy number $p$-values are much smaller than the smallest gene
expression $p$-values. Thus, application of the B--H procedure to copy
number $p$-values yields 62 significant genes, while application to the
expression $p$-values fails to yield any. It is unsurprising then that
application of the B--H procedure to the set of all maximum $p$-values for
each gene also produces no significant results.

%
Using the Voronoi $p$-value combination followed by the B--H procedure on
the summarized values at $\alpha=0.05$ gives 12, 14 and 25 rejections
for Euclidean, Maximum and Summation orderings. Guided by the results
of simulation, we consider the rejections made using the Summation
ordering. Of these 25, four were mapped to official gene names, and all
four were listed in the COSMIC database of cancer genes [\citet
{forbes11}]. These four genes are CABLES2, PAK1IP1, CAMKV and TSHZ1.
The COSMIC results suggest that there are mutations in these four genes
that are found in a variety of cancers, thus strengthening the evidence
of these genes being putative oncogenes in prostate cancer.

To use DAVID [\citeauthor{da08} (\citeyear{da08,da09})] for further investigation of our
results, a larger gene list was necessary. For this purpose, we
performed the B--H procedure on the combined $p$-values at $\alpha=0.20$.
Under the summation ordering, this yielded 306 rejections, 102 of which
could be mapped to recognized genes by DAVID. The functional annotation
tool found significant enrichment (adjusted $p$-value of 0.019) in the
Fibrinolysis pathway. Fibrinolysis has\vadjust{\goodbreak} been associated with prostate
cancer for decades [\citet{tagnon52}]. Tumor classifications for
different malignancies have been proposed based on the behavior of this
pathway [\citet{zacharski92}]. Results for functional classification of
the 102 genes are summarized in Table~\ref{tab:davidfunctional}.

\begin{table}
\tabcolsep=0pt
\caption{Summary of Gene Functional Classification from DAVID}
\label{tab:davidfunctional}
\begin{tabular*}{\textwidth}{@{\extracolsep{\fill}}lcc@{}}
\hline
\multicolumn{1}{@{}l}{\textbf{Number}}&\textbf{Enrichment}&\\
\textbf{of genes} & \textbf{score} & \multicolumn{1}{c@{}}{\textbf{Keywords}}\\
\hline
16& 0.73 & Peptidase; Serine; Endopeptidase; Kringle\\
11 & 0.58 & Transmembrane; Membrane; Extracellular; Cytoplasmic\\
\phantom{0}3 & 0.50 & Transport: protein, intracellular, vesicle; Golgi apparatus\\
\phantom{0}3 & 0.36 & Catabolic process; Proteosome; Proteolysis\\
12 & 0.27 & Lumen: nuclear, intracellular, organelle,
membrane-enclosed;\\
& & Phosphoroprotein; nucleolus; ATP binding\\
\phantom{0}3 & 0.22 & GTP-binding; Nucleotide binding: guanyl, purine;\\
& & Ribonucleotide binding: guanyl, purine\\
20 & 0.12 & Transmembrane; Membrane; Glycoprotein; \\
\hline
\end{tabular*}
\end{table}
%
\section{Discussion}
\label{sec:discussion}
In this paper we have presented a novel approach to $p$-value combination
for testing the disjunction hypothesis when two $p$-values are considered
for each test. The approach uses an extension of one-dimensional
spacings, Voronoi cell areas, in combination with concave ordering
schemes to define cumulative areas usitable for multiple testing
techniques. When the majority of $p$-vectors have independent components,
techniques such as the B--H procedure can be directly applied. If the
components are correlated, empirical null techniques are more suitable.
Simulation studies showed that the approach has appropriate error
control properties and results in a gain of power over the existing
method. This increased power is of particular interest for detection of
genes related to biological processes or implicated in cancer progression.

Four candidate ordering schemes were described, and simulations were
used to test their performance in several settings. The concave up
ordering proposed by \citet{lichtenberg05} failed to control FDR in the
paradigm of the disjunction hypothesis. As discussed in Section~\ref{subsec:independence}, we suspect that concavity of an ordering's
contour lines is vital to its FDR control characteristics.
Specifically, as contours become increasingly concave down, the
procedure is more conservative. The reverse applies when considering
concave up schemes. For this reason, we recommend using the summation
ordering in practice, as it represents the boundary case between
concave up and down. This offers the least conservative, and thus most
powerful, procedure that retains appropriate FDR control.

This approach can be extended in several meaningful directions. The
conjunction or partial conjunction hypotheses could be tested by
defining suitable ordering schemes such as the minimum, or product.
Extension to higher dimensions is also of utmost interest, particularly
considering the scale of current biological and genomic experiments. In
Section~\ref{sec:extension} we described a potential extension to three
or more dimensions, but further investigation of this and other
techniques is necessary.

\section*{Acknowledgments}
The authors would like to thank the area
Editor and an anonymous referee whose comments greatly improved the
quality of this paper.


\begin{supplement}[id=suppA]
\sname{Supplement A}
\stitle{Summarized results of additional simulation
studies\\}
\slink[doi]{10.1214/13-AOAS707SUPPA} 
\sdatatype{.pdf}
\sfilename{aoas707\_supp.pdf}
\sdescription{We present the summarized results of the proposed procedure's
performance in the challenging situations described in Section~\ref{sec:depsims}. Results include estimated FDR and 1-NDR for each of the
two settings.}
\end{supplement}

\begin{supplement}[id=suppA]
\sname{Supplement B}
\stitle{Supplementary code and data\\}
\slink[doi,text={10.1214/13-\break AOAS707SUPPB}]{10.1214/13-AOAS707SUPPB} 
\sdatatype{.zip}
\sfilename{aoas707\_supp.zip}
\sdescription{R code including
the functions required to perform the procedure described in this
paper, to replicate the described simulation studies and to perform the
described data analysis. The relevant data sets are also included.}
\end{supplement}

%

%

\printaddresses

\end{document}